\documentstyle[12pt,aaspp4]{article}

\begin{document}
\title{An Einstein X-ray Survey of Optically-Selected Galaxies: I. Data}
\author{David Burstein}
\affil{Dept. of Physics \& Astronomy, Arizona State University, Tempe, AZ\\
Email: burstein@samuri.la.asu.edu}
\author{C. Jones, W. Forman}
\affil{Harvard-Smithsonian Center for Astrophysics, Garden St., Cambridge, MA
02138\\Email: cjf@cfa236.harvard.edu, wrf@cfa.harvard.edu}
\author{A. P. Marston}
\affil{Dept. of Physics \& Astronomy, Drake University, Des Moines, IA 50311\\
Email: tm9991r@acad.drake.edu}
\author{Ronald O. Marzke}
\affil{Harvard-Smithsonian Center for Astrophysics, Garden St., Cambridge, MA
02138\\
and\\
Dominion Astrophysical Observatory, Herzberg Institute of Astrophysics\\
National Research Council of Canada\\
5071 W. Saanich Rd., Victoria, BC, Canada V8X 4M6\\
Email: marzke@dao.nrc.ca}

\abstract

We present the results of a complete Einstein Imaging Proportional Counter
(IPC) X-ray survey of optically-selected galaxies from Shapley-Ames Catalog,
Uppsala General Catalog and the European Southern Observatory Catalog. 
Well-defined optical criteria are used to select the galaxies, and X-ray
fluxes are measured at the optically-defined positions. The result is a
comprehensive list of X-ray detection and upper-limit measurements for 1018
galaxies.  Of these, 827 have either independent distance estimates or radial
velocities.  Associated optical, redshift and distance data has been assembled
for these galaxies, and their distances come from a combination of directly
predicted distances and those predicted from the Faber-Burstein Great
Attractor/Virgocentric infall model.

The accuracy of the X-ray fluxes has been checked in three different ways; all
are consistent with the derived X-ray fluxes being of $\le$0.1 dex accuracy.
In particular, there is agreement with previous published X-ray fluxes for
galaxies in common with Roberts et al. (1991) and Fabbiano et al. (1992). The
data presented here will be used in further studies to characterize the X-ray
output of galaxies of various morphological types and thus to enable the
determination of the major sources contributing to the X-ray emission from
galaxies.

\section{Introduction}

Almost all X-ray surveys of galaxies published to date have used galaxy samples
chosen on the basis of apparent optical brightness.  These include the first
surveys of the HEAO-2 (Einstein) data base (Long \& Van Speybroeck, 1981;
Forman, Jones \& Tucker, 1985; Fabbiano \& Trinchieri, 1987; Canizares,
Fabbiano \& Trinchieri, 1987, and references therein; Roberts et al. 1991), as
well as the compilation by Fabbiano et al. (1992; hereafter F92).  In total 405
galaxies have Einstein measurements of X-ray properties studied in these
surveys, all chosen from either the Shapley-Ames Catalog (Shapley \& Ames
1932; hereafter S-A) or the Second Catalog of Bright Galaxies (de Vaucouleurs,
de Vaucouleurs \& Corwin 1976; here after RC2).  As analyzed by F92, 207 of
these galaxies have 3-$\sigma$ X-ray detections.

However, consider the analogy with stars:  The H-R diagram of the 100
brightest stars in the sky is very different from that of the 100 nearest
stars.  Is it possible that the same kinds of differences occur in galaxy
samples similarly chosen, with the additional option that galaxies can be
selected either by apparent brightness or by apparent size?  A direct answer 
to this question is needed for a physical property of galaxies such as X-ray 
flux.  The X-ray flux can come from several different sources within a 
given galaxy (e.g., hot thermal gas; accretion disks around binaries; 
central engines of active galaxies).

Given this motivation, the observational goal of the present survey was
to obtain measurements of the X-ray flux (or upper limits) at the positions of 
galaxies optically-selected in two different ways, as viewed in the Imaging 
Proportional Counter (IPC) fields obtained by the Einstein Observatory.  One
set, from the S-A catalog, is selected on the basis of apparent magnitude.
Two data sets, one from the Uppsala General Catalog (Nilson 1973; hereafter
UGC) and Europeran Southern Observatory catalog (Lauberts 1982; hereafter ESO)
are chosen on the basis of apparent size.

In the present paper, we present the X-ray measurements we made and the
associated optical, redshift and distance data for 1018 galaxies. A second
paper will analyze the global properties of galaxies as found by this survey.
\S~2 discusses how the galaxies were selected for study, and how the X-ray
data were obtained. In \S~3 we discuss the origin of the non-X-ray data used
for the galaxies in this survey --- radial velocities, distances, magnitudes
and diameters.  Many of the normal galaxies with published X-ray fluxes from
Einstein fields have had their X-ray fluxes independently determined by this
survey, in order to permit ready correspondence to previous results.  Tests of
the uniformity of the X-ray data are detailed in \S~4, including a detailed
comparison with the F92 published X-ray counts and fluxes.  \S~5 gives a
summary of the paper.

\section{Galaxy Selection and Einstein X-ray Observations}

\subsection{Combined Optical/X-ray Selection of Galaxies}

Most tests of the inclination-dependence of the observed properties of galaxies
show that while the magnitudes of E and S0 galaxies are inclination-invariant,
those of spiral galaxies are inclination-dependent (cf. Burstein, Haynes \&
Faber 1991).  This difference is due to the fact that spiral galaxies have 
visible dust distributed in their disks (cf. Giovanelli 1995), while the dust 
seen in E and S0 galaxies is well-known to be localized near their centers. 
Specific tests on the UGC-measured and ESO-measured diameters of spiral
galaxies indicate little change with inclination (e.g. Valentijn 1990;
Burstein et al. 1991).  However, interpreting these tests in terms of whether
spiral galaxies actually have inclination-independent isophotal diameters is
still in debate (cf. Davies, Jones \& Trewhella 1995; Burstein, Willick \&
Courteau 1995; Giovanelli 1995).

Nevertheless, it is clear that the inclination-dependencies of magnitudes and
diameters for E/S0's is different from that for spiral and irregular galaxies.
As such, it is not surprising that their relative numbers are different in a
magnitude-limited sample (spirals undersampled), relative to a
diameter-limited sample (spirals oversampled).  One finds that galaxies of
types E, S0 and S0/a comprise 24\% of the S-A catalog (Sandage \& Tammann
1981), while such galaxy types comprise only 16\% of either the UGC or ESO
catalogs. The present X-ray survey includes galaxies from all three
catalogs, as one check for inclination-dependent selection effects influencing
the results of X-ray studies of galaxies.  Taken together, the data from all
three catalogs comprises the most complete survey one can do for the brightest
galaxies with the Einstein data base.

The methodology we employ is to use the optical positions of galaxies
to locate apertures of known size at the corresponding locations in the
Einstein images, and then to measure the observed count rate within these
apertures, properly correcting for background count rate.  As we expect
to find upper limits to the fluxes of most galaxies, the need for accurate
detection statistics make the use of relatively long exposures desirable. 
A lower exposure limit of 2000 seconds was chosen for images obtained with the 
Imaging Proportional Counter (IPC) for UGC and ESO galaxies, and a lower limit 
of 1500 seconds was used for S-A galaxies (owing to the fact that the S-A 
galaxies are, in general, closer and generally of higher flux than the UGC and 
ESO galaxies).  After some experimentation, High Resolution Imager (HRI) 
exposures were not used for the final sample as the lower sensitivity of the 
HRI yielded much higher signal-to-noise upper limits, on average.

Galaxy selection for the present samples proceeded by cross-correlating the 
galaxy positions in each catalog with the positions of all Einstein IPC fields 
with at least the minimum exposure time.  Each IPC pixel is 8 arc-sec on a 
side, hence galaxy positions accurate to $\approx$10 arc-sec are sufficient 
for the purposes of this survey.  Galaxies in the ESO and S-A catalogs have 
positions generally this accurate, and their quoted positions were used 
directly in the cross-correlation with IPC image positions.

Galaxies in the UGC have positions of mixed accuracy, some of which are very
poor ($>$1 arc-minute).  Improved positions for UGC galaxies were obtained in 
a two-step process:  First, a cross-correlation was done between UGC-listed 
galaxy positions and IPC field positions.  After further objective selection 
criteria were applied (common to all three catalogs and discussed below), a 
final list of 773 UGC galaxies was compiled.  Second, accurate positions were 
either obtained from Dressel \& Condon (1976) (320 of their 1800 galaxies are
in our sample), or were measured by one of us (ROM) in the following manner:
Transparent overlays were made of the field around the UGC position of each 
galaxy, to the scale of the Palomar Sky Survey (PSS) prints.  Star positions 
and the UGC galaxy position were marked, and the direction of N-S and E-W 
drawn.  This overlay was then placed on the PSS print, and the position of the 
galaxy on the print, relative to the UGC position, was measured with a 
graduated reticle in both Right Ascension and Declination.  Comparison with 
35 positions measured by Dressel \& Condon indicate that the positions for 454
galaxies measured in this manner are accurate to $\pm$12 arc-seconds, 
adequate for the purposes of this survey.  A post-survey re-comparison with
the UGC indicates that we missed at most 15 galaxies (for reasons such as
bad search positions; IPC image not yet processed, etc.) that could 
otherwise have been in our survey.  This omission has a negligible effect
on the results of this survey.

Three optical selection criteria are applied to all three catalogs: galaxies 
with galactic reddenings, E(B-V) $>$ 1.00 (from Burstein \& Heiles 1984 and
unpublished) were excluded, as well as galaxies without a listed morphological
type.  Galaxies with listed optical diameters larger than 5$'$ are excluded
from the UGC and ESO catalogs, but not necessarily from the S-A catalog.
Multiple galaxies listed in the UGC and ESO catalogs are included in
this survey.  X-ray selection criteria are discussed below.

The combination of optical and X-ray selection criteria results
in 773 UGC galaxies in 483 IPC fields, 757 ESO galaxies in 232 IPC fields and
404 S-A galaxies in 374 IPC fields. Of the 773 UGC galaxies, 103 are also
in the S-A catalog; of the 757 ESO galaxies, 72 are in the S-A catalog. This
survey therefore includes a total of 1759 galaxies with 2147 galaxy 
observations in 942 IPC fields (97 of these galaxies had X-ray counts 
measured by more than one of us independently).  In addition, 194 measurements
are taken at random positions in 8\% of the IPC fields (see \S~2.2).

\subsection{X-Ray Flux Determinations}
\label{HIcor}

In obtaining the raw counts for the source and background regions from the IPC
images, we used the energy range from 0.56 to 4.47 keV. Four box apertures are
employed for measuring source counts, having sizes of 200\arcsec, 232\arcsec,
264\arcsec ~and 296\arcsec ~on a side.  The size of the largest aperture was
dictated by the necessity of uniformly measuring many galaxies in arbitrary
positions on the IPC images.

The X-ray source and background counts on the IPC images were visually
examined using the HEAO Image Processing monitor.  Each IPC image was
displayed and the position of each galaxy marked on the field.  As with
photoelectric aperture photometry of galaxies, in measuring background count
rate, care was taken to avoid regions where the background was high, varying
spatially, or where the IPC image was obstructed by the detector window support
structure ('rib').

If the galaxy position lay close to a 'rib', or close to the edge of the field,
the smaller apertures were measured if they could yield good data. However, in
the majority of cases, placement near a 'rib' or near the edge of the field
meant that counts could not be measured for the galaxy.  If the galaxy
position lay within a region of diffuse emission (such as exists in a galaxy
cluster), the galaxy was similarly excluded from X-ray measurement. Table~1
gives the numbers of galaxies rejected and the reasons for rejection.
Background count rates were measured for each galaxy, both near the galaxy
location and at 'mirror-image' positions on the same field, after confirming
that the background positions did not contain any obvious X-ray sources.

Two internal checks of our measuring procedure are made during the course 
of this survey.  First, measurements are made at the positions of 194 'blank'
fields, using the same procedures as for the sample galaxies. These fields
are chosen in 73 IPC images that sample the range of exposure times used in
this survey.  The data for these 'blank' fields are then processed in the same
manner as for the galaxies.  Second, due to the labor-intensive, interactive
manner of deriving the aperture measurements, it is desirable that a
subset of the galaxies be measured twice, once each by different ``observers''.
The IPC emission of 97 galaxies were measured twice, once during the UGC/ESO
phase of the program and once later during the S-A phase.  Comparison of the
two sets of observations yield very close agreement.  In order to preserve
similar signal-to-noise, we do not average these two separate measurements of
the same IPC image, but rather use just one.

X-ray fluxes are computed from the IPC images by correcting the net source
counts in each aperture measured for the following explicit issues:  a) the
exposure time; b) the instrument deadtime ($\sim$4\%); c) telescope
vignetting (Harris {\it et al.} 1991); d) HI column density;  e) spectral
energy distribution of the source; and f) predicted ratio of flux within 
measured aperture to total flux.  The first three corrections are well-known;
for the vignetting corrections we assume a mean photon energy of 1.49 keV.

The next two corrections require knowledge of the HI column density (taken
from Stark et al. 1992 and sources given in Marshall \& Clark 1984) and the
spectral energy distribution of the source.  As the total number of counts for
even the detected galaxies in our sample is generally too small to constrain
the energy spectrum, we must assume a spectral energy distribution for these
galaxies.  As a test, we compare the differences in conversion rates that
would arise from three different forms of the spectrum: a) a 1 keV Raymond
spectrum as appropriate for luminous ellipticals; b) a 7 keV exponential
spectrum as characteristic of binary X-ray sources; and c) a power law
spectrum with a photon index of 1.5 typical of AGN.  For each kind of spectrum
we generate the count rate to flux conversion for a range of HI values from $8
\times 10^{19}$ to $3 \times 10^{21}$ $\rm N_H \, cm^{-2}$.  The hard
exponential and power law spectra give nearly the same count rate to flux
conversions.  The ratio of the 1 keV thermal conversion value to the other
conversions ranges from 0.75 for low HI column density to 0.85 for high HI.

However, the X-ray emission mechanism for each galaxy, or even each type of 
galaxy is not certain.  For example, although binary sources are found in 
spiral galaxies, at least some spirals also have substantial diffuse emission.
While hot gas seems to dominate the X-ray emission from ellipticals, many
X-ray detected early-type galaxies contain active galaxy nuclei.  Thus, at this 
stage in our analysis, we are reluctant to assign a particular spectrum to a 
particular class of galaxy and thereby bias the flux conversions.  Instead we 
choose to use an average count rate to flux conversions dependent only on the
galactic HI column density, as given in Table~2 for all the galaxies in our 
samples, which is a compromise among the results from the three models.

The aperture-related correction is necessary as some galaxies could have
fluxes only determined in smaller apertures.  Smaller apertures are used when
a source is near the edge of the field, near a rib, or near another source.
Hence, data from all four X-ray apertures are used to examine the spatial
distribution of the X-ray sources that are detected.  Growth curves calculated
from bright X-ray sources were used to extrapolate the flux from the largest
aperture available for each galaxy observation, to a total flux.  These
corrections, listed in Table~3, are computed separately for elliptical
galaxies, spiral galaxies and central point source-dominated galaxies (such as
Seyfert galaxies).  Table~4 gives the number of galaxy observations for which
each aperture was the largest measured.  X-ray fluxes for 82\% of the UGC
galaxies with X-ray observations, 75\% of the ESO galaxies, and 94\% of the S-A
galaxies are measured in the largest aperture used in this survey 
(296\arcsec ~on a side).  For those galaxies whose measured aperture is a 
smaller aperture, the observed flux in that smaller aperture is corrected 
to that of the largest aperture using the ratios in Table 3.

Signal-to-noise is calculated in the usual manner assuming a Gaussian
distribution of errors and taking into account both galaxy and background
measurements.  From our accuracy tests (see below), a signal-to-noise ratio of
2.5 was chosen to distinguish between X-ray detections and non-detections in
this sample. If a galaxy has S/N$\ge$2.5, the flux given is the flux observed.
If a galaxy has S/N $<$ 2.5, the flux given is 2.5xN.

Finally, it should be clear that, although the selection of galaxies was done
in an objective manner, only 5\% of the sky is covered by IPC images.  As many
galaxies are IPC targets, the net effect is to incorporate a combination of
several selection effects.  We keep track of this effect by the size of the
vignetting correction, which is close to unity for objects near the center
of the Einstein IPC images.  In the cases for galaxies observed on more than 
one IPC image, the vignetting correction value closest to unity are kept.

The number of galaxies included in this survey, relative to the total number
of galaxies in the relevant optical catalogs is shown in Figure~1. Galaxies are
divided into four sections by morphological class: ellipticals and S0's,
early-type spirals, late-type spirals and irregulars, and others (e.g. dwarfs,
peculiars, multiple galaxies). This last category was used only for the UGC
and ESO catalogs.  Within each class, in Figure~1 we plot the fraction of
galaxies included in the X-ray sample compared to those available from the
whole catalog as a function of either apparent size (UGC and ESO) or apparent
magnitude (S-A).

As is evident, the percentage of galaxies in the S-A sample with measured X-ray
fluxes is an order of magnitude higher than in the diameter-selected sample.
Yet, due to the order of magnitude higher number of galaxies included in
the UGC and ESO catalogs, the total number of galaxies with measured X-ray
fluxes is comparable in each catalog.  Within statistical fluctuations, the
X-ray observed samples are fair cross-sections of both morphological
types and apparent size (or magnitude) for these three catalogs.

\section{Optical, Redshift and Distance Data}

\subsection{Optical Data}

Optical diameters, magnitudes and axial ratios for blue magnitude data are
given for each galaxy in this survey when available.  The primary source for
the S-A sample is the RC2, for which apparent magnitudes (either on the Harvard
corrected system, or on the BT system) are given, together with isophotal
diameters (to the 25th mag isophote) and axial ratios. UGC galaxies have
optical data obtained primarily from the UGC, for which diameters, axial
ratios and magnitudes are listed.  UGC galaxy magnitudes that are 15.7 or
brighter come from the Zwicky {\it et al} (1961) catalog.  Zwicky magnitudes
require correction to be placed on a common system in reasonable accord with
the RC2 system.  The correction procedure adopted here is that advocated by
Huchra (1976).

ESO galaxies have optical data obtained from the more recent photographic 
surface photometry published by Lauberts \& Valentijn (1988). This photometry 
is available for 85\% of the original 16,154 galaxies in the ESO survey and,
as such, is available for a similar percentage of the galaxies in the 
present survey. The 25th blue magnitude isophotal diameters and total
blue magnitudes used here are taken from Lauberts \& Valentijn; axial ratios
are taken from the original ESO catalog (Lauberts 1982).

Galactic reddenings, E(B-V) are taken from Burstein \& Heiles (1984) for the
RC2 and UGC galaxies, and separately calculated using the Burstein-Heiles
reddening maps for ESO galaxies.  The reddenings are expressed in terms of
Burstein-Heiles-defined galactic extinction, $A_B = 4\times E(B-V)$. 
Redshift-dependent K-corrections are calculated using the precepts of the RC2.

\subsection{Redshifts and Distances}

Heliocentric radial velocities are obtained primarily from the
computer-readable version of the Third Reference Catalog of Bright Galaxies
(de Vaucouleurs et al. 1991; hereafter RC3), as distributed by H.G. Corwin, Jr.
These values are then supplemented by examination of the NASA Extragalactic
Database (NED), as of February, 1996.

Distances for galaxies are determined in a two-step process. First, all of the
galaxies in this sample are cross-correlated with the galaxies in the Mark~III
Catalog of Galaxy Peculiar Velocities (Willick et al. 1995; 1996a,b;
references cited in notes to Table~7).  We use the distances termed
``inhomogeneous Malmquist-corrected" as given in the Mark~III. Distances of
galaxies are expressed in km sec$^{-1}$ units, using $\rm H_0 = 75$ km
sec$^{-1}$ Mpc$^{-1}$.  Galaxies without directly measured distances, but with
measured radial velocities have their distances calculated, in terms of km
sec$^{-1}$, by the Great Attractor velocity field model of Faber \& Burstein
(1988).

\subsection{The Observed and Derived Data Sets}

Tables 5a, b and c contain the observed data for this survey for the galaxies
in the UGC, ESO and S-A catalogs, respectively.  Only those galaxies with X-ray
measurements are included. This includes 313 S-A galaxies, 393 UGC (not S-A)
galaxies and 312 ESO (not S-A) galaxies, for a total of 1018 galaxies with
X-ray observations.  These have the following format: Column (1) gives the
name of the galaxy; column (2) the apparent blue magnitude of the galaxy;
column (3) the logarithm of the blue isophotal diameter; column (4) the 
morphological type number, T (as explained in the notes to this Table).
Column (5) gives the Einstein field number used for the X-ray data.  
A value of -1 indicates that multiple fields were used (see Table~8 for a 
listing of these fields).  Column (6) gives the total exposure time for
this galaxy (including all IPC images sampled).  Columns (7) and (8) give the 
RA (1950) and Dec (1950) position at which the X-ray image was searched.  
Column (9) gives the X-ray count rate measured for this galaxy, in units of 
1000 times the observed count rate. This count rate is only corrected for 
background and deadtime correction. Column (10) gives the error on this count 
rate, in the same units.  The values in Column (11) are the multipliative 
IPC count rate to flux conversion for the energy range from 0.5 to 4.5 keV 
for an IPC count rate of 1 count/second and an X-ray spectrum, as discussed in 
section \S \ref{HIcor}.  Column (12) contains the IPC vignetting correction.  
If more than one IPC image is used for the observed count rate, the vignetting 
correction given here is the exposure-weighted average of the individual 
vignetting corrections. Column (13) indicates which of the four apertures were 
used to obtain the observed count rate (1 = smallest, 4 = largest).

Tables 6a, b and c list those galaxies included in the original survey, but
for which X-ray measurements were not made. The reasons for exclusion
(discussed in more detail previously in \S~\ref{HIcor}) include obscuration by
the `ribs' of the IPC; near or off the edge of the IPC image; only in HRI
images; or within diffuse X-ray emission.  Eleven S-A galaxies that were too
large for the apertures used in this survey also were not observed. In
Tables~6, column (1) has the name of the galaxy; column (2) the apparent blue
magnitude of the galaxy (if available); and column (3) the numerical code
indicating the reason for omission (see notes to the tables for an 
explanation of this code.)

Tables 7a,b,c contain the distance-dependent derived parameters for the X-ray
measured galaxies in this survey.  Column (1) gives the galaxy name; column
(2) the heliocentric radial velocity.  Column (3) gives the distance used for
this galaxy, in units of Mpc (assuming $H_0$ = 75 km sec$^{-1}$ Mpc$^{-1}$),
while column (4) gives a numerical code for the source of this distance, as
detailed in the table notes.  Columns (5), (6) and (7) give the absolute B
magnitude, log of galaxy diameter (in kpc) and fully-corrected absolute X-ray
luminosity for this galaxy.  Column (8) notes which galaxies are generally
classified as Seyfert galaxies.  Tables 7 include 329 UGC galaxies, 180 ESO
galaxies and all 313 S-A galaxies, for a total of 822 galaxies (81\% of the
total sample) with both X-ray observations and known distances. The 196
galaxies in the UGC and ESO samples that have measured X-ray fluxes (both
detected and upper limits), but do not yet have radial velocity measurements
are given in Table~9.  Included among these are 11 galaxies whose X-ray
emission was detected with Einstein.

\section{Internal and External Estimates of the Accuracy of the X-ray Data}

\subsection{Sky Measurements and 'Blank' Fields}

For 251 of the 1018 galaxies with X-ray measurements (24.7\%), the observed
X-ray flux has a signal-to-noise (S/N) of 2.5 or better and a redshift is
known.  Correcting this percentage for the number of false detections, which
we determined through a similar analysis of "blank" fields, we have a
detection rate of 22.6\%.  Most of the galaxies in this survey have only X-ray
upper limit values.  The fact that most galaxies in this survey would not be
detected at a significant level was expected at the outset.  Indeed, one of
the principal aims of this survey was to assemble enough galaxies in each
morphological class so that an average X-ray flux per class could be derived,
even if most individual galaxies were not detected.  Given this intent, it was
important to be able to verify the statistical accuracy of noise estimates.

This was done in two ways. First, the 194 blank fields were measured. As noted
above, these were treated in the same way as the galaxy measurements.  S/N
histograms from the blank field observations were compared to those formed
from the program galaxies.  Second, comparisons were made among the 97 galaxies
which were twice measured independently.

Figure~2 shows the histogram of the S/N measurements of the blank field
measurements. Also plotted in this histogram is the expected Gaussian
distribution of S/N.  There is a slight tendency for the blank fields to show
a small net positive S/N (+0.095) compared to a Gaussian distribution.
Nevertheless, the measurement of these blank fields is not expected to
produce a Gaussian distribution based at S/N = 0, but rather be representative 
of what we would have measured in the absence of X-ray flux at these galaxy
positions.

In Figures~3 and 4 we show X-ray measurement signal-to-noise distributions for
E/S0 galaxies (Figure~3) and spiral and irregular galaxies (Figure~4) separately
for the three catalog samples, and for all of the data together.  Plotted
over these histogram distributions is the measured distribution for the blank
fields, scaled to match the number of galaxies actually measured in each
sample.  It is satisfying that the negative S/N tail of the blank fields
defines the minimum S/N of all of the galaxy samples in which most galaxies
are not detected.  This is particularly true for spiral and irregular galaxies,
and gives us confidence in the accuracy of our measurements.

The histograms in Figures~3 and 4 show two trends.  First, it is obvious
that almost all of the S-A galaxies have detected X-ray flux at some
level.  In contrast, for most of the non-S-A UGC and ESO galaxies, the measured
X-ray fluxes are clearly upper limits.  Second, in each catalog it is evident
that more E/S0 galaxies are detected than are spiral/irregular galaxies.  
Indeed, among the spiral/irregular galaxies, almost all of the X-ray dectected 
galaxies are in the S-A; very few are in the UGC or ESO samples.

Our choice of S/N = 2.5 as the dividing line between detected and 
undetected galaxies in our survey comes from two results evident in these
histograms.  First, only 4/194 = 2\% of the blank fields have S/N values
greater than 2.5.  Second, the negative S/N tails of the blank fields and of
the spiral/irregular UGC and ESO galaxy samples are well matched, indicating
the galaxy observations have a similar intrinsic S/N distribution as the
blank fields.  Based on the blank field observations, we would estimate that
as many as 5 galaxies with fluxes near the S/N = 2.5 cutoff (2\% of the 
detected sample) could have spurious detections.

\subsection{Comparison to the Published Data}

Fabbiano et al. (1992) have published Einstein observations for 405 individual
galaxies, all of which are taken from the RC2 catalog.  Given the format in
which their data are published, we can compare the results of our survey with
theirs in two ways:  by count rate and by fully-corrected X-ray luminosity.
For this comparison we find that we should exclude 22 galaxies from F92 that
have only HRI observations, as we have used only IPC observations for our
samples.

Upon further comparison, we find only 285 of the remaining 383 galaxies 
have quoted count rates and luminosities in common between both samples.  
Of the 98 galaxies that have F92 IPC observations but are not in our sample,
almost all were excluded from our sample due to various objective selection 
criteria (e.g., size, diffuse emission, non-UGC/ESO/S-A, too short an image
exposure).  Ten galaxies are not in this comparison as they are among the few
UGC galaxies not observed for the present survey.

In 112 of these 285 galaxies in common, both our data and those of F92 are 
upper limits.  As comparing just the upper limit values for galaxy fluxes is 
inconclusive, this leaves us with 173 galaxies for which we can make a 
meaningful comparison with F92.  Of these, 137 have detections in both data 
sets, 25 are detected by us and not by F92 and 11 are detected by F92 and not 
by us.  Our survey has detected 251 galaxies (including those galaxies with 
and without radial velocities) at the 2.5$\sigma$ level or higher.  Thus,
our sample contains 89 X-ray detected galaxies not previously published in the 
F92 compilation, a 43\% increase over the 207 X-ray detected galaxies listed 
by F92.

That we have generally good agreement between our fully-corrected X-ray
luminosities and those of F92 is shown in Figure~5 for the 137 galaxies
detected in common between the two data sets.  F92 fluxes have been adjusted
to our distances for this comparison.  As can be seen from Figure~5, while the
relative flux scales between F92 fluxes and ours are the same, there are some
differences in zero points. To examine this latter point more closely,
Figure~6a shows the logarithm of the ratio of the F92 X-ray fluxes to our
fluxes, plotted versus our X-ray fluxes, for galaxies detected by either
data sample.  The two dotted lines give the median values for 
$\rm \log [L_X(US)/L_X(F92)]$ as a function of Hubble type only
for the galaxies detected by both surveys: -0.05 for E, S0 and I0 galaxies; 
-0.12 for spiral galaxies.

Such Hubble type--dependent differences can arise either from differences in
energy range observed, or in HI-dependent correction for spectral energy
distribution differences. To separate these two effects, in Figure~6b we plot
the logarithm of the ratio of the observed count rates, $\rm \log
[Ct(US)/Ct(F92)]$ as a function of Hubble type, using the RC3 numerical code
for Hubble type (see notes to Table~5a).  The four galaxies detected in this
survey with fluxes and countrates that differ signficantly from those of F92
are noted by NGC number.  In Figure~6b the dotted lines correspond to the
median log ratios of count rates, Us to F92, of -0.12 for E, S0 and I0 
galaxies and -0.10 for S0 and spiral galaxies.  To an accuracy of 0.01 dex,
our count rates differ from those of F92 by 0.1 dex, in the sense of our
count rates being lower.

We thus see that the differences between our fluxes and those of F92 are a
combination of differences in observed count rates and differences in
conversion of counts to flux.  That our observed count rate is systematically
lower than that of F92 is to be expected as we used a different energy range
with the IPC than they did: F92 is quoted as using 0.2 to 4.0 keV; we use 0.5
to 4.5 keV. Similarly, it is also the case that our final luminosities should
differ somewhat from those quoted by F92 as well.  F92 used a bremsstrahlung
spectrum of 5 keV for the spirals and irregulars and a Raymond spectrum of 1
keV for the ellipticals and S0's.  As explained previously (\S \ref{HIcor}),
for our sample we did not {\it a priori} distinguish among Hubble types as to
the physical source(s) of their X-ray spectra in converting observed countrate
to X-ray flux.

Of the four galaxies which show significantly different F92 fluxes than our 
measurements, NGC~2832, NGC~1399 and NGC~4406 are all elliptical galaxies 
whose measured X-ray flux can be influenced by emission from diffuse cluster 
gas.  NGC~5850 is a spiral galaxy with relatively low count rate (0.0053 
cts/sec).  The sense of the difference, Us/F92 is that in the case of the 
three elliptical galaxies, our measurements do not include any diffuse cluster 
gas, while in the case of the spiral galaxy, it is likely due to a small
difference in how the background was treated.

Given the current uncertainty in the true X-ray energy distributions from
galaxies, we accept that differences of $\sim 0.1$ dex can occur among quoted
luminosities derived from the same data under different assumptions. Indeed,
we find it comforting that, to a level of 0.1 dex, the X-ray fluxes measured
by F92 and by us are consistent with each other.  While we find 6 galaxies
(3.5\% of galaxies detected by one or the other survey) have signficant
discrepant fluxes between the two data samples (including 2 upper limits),
this is to be expected.  Such differences can occur due to such problematic
measurement issues as background flux estimation on the IPC images and
possible confusion of cluster diffuse emission with galaxy emission.  In both
cases, our measurement technique differs enough from that used by F92 to lead
to possible large differences in a few cases.

We believe our method of doing essentially X-ray aperture photometry with
the Einstein data base is complimentary to the technique employed by F92.
In that paper, following their previous studies, those authors dealt with
the IPC and HRI images in an analogous manner as one would do CCD surface 
photometry of galaxies, including global mapping of the response function.  
In contrast, our technique is more akin to photoelectric aperture photometry, 
complete with determining the X-ray ``sky background'' separately for each 
object.

Checks also were made of our measurements against earlier compilations (e.g.
Forman, Jones, \& Tucker 1985; Roberts et al.) and very good consistency was
found. However, for these comparisons we note that the same procedures, by a
subset of the same authors, were used to derive the fluxes.
                         
\section{Summary}

We have made an X-ray survey of the Einstein IPC images which contain galaxies
found in three catalogs --- the Shapley-Ames, the UGC and the ESO catalogs ---
selected acoording to well-defined objective criteria.  The X-ray fluxes for a
total of 1018 galaxies were obtained, 313 in the S-A catalog, 393 non--S-A
galaxies in the UGC and 312 non--S-A ESO galaxies.  Our manner of obtaining the
X-ray fluxes is analagous to obtaining optical photoelectric measurements:  We
locate apertures of defined sizes at the optical positions of the galaxies,
and sample background at symmetric areas of the IPC field.  In this way, our
analysis differs from that done by Fabbiano et al. (F92), for which the
analogy of CCD-like surface photometry is more applicable.

For these galaxies we have assembled associated optical data and have assigned
distances either according to direct measurements (Tully-Fisher for spirals
and $\rm D_n-\sigma$ for ellipticals) or to the velocity field model of Faber
\& Burstein (1988).

The accuracy of our X-ray fluxes has been checked in four independent ways.
First, we also measured X-ray flux at 194 random IPC positions in the same
manner as for our program galaxies.  The signal-to-noise histogram of these
``blank'' fields is reasonably Gaussian in distribution, with an offset
(+0.095) in S/N and only four observations lying beyond the 2.5$\sigma$
level.  Hence, we choose the lower limit for an X-ray detection in this survey
to be 2.5$\sigma$.  Second, X-ray fluxes on 97 IPC images were measured
independently by two of us.  Third, the negative side of the S/N histogram of
the blank fields matches well the negative side of the S/N histogram of the
program galaxies with the least intrinsic X-ray flux --- the spiral and
irregular galaxies.

Fourth, we have made a quantitative comparison with the X-ray fluxes and
count rates published by Fabbiano et al. for 173 galaxies in common for
which a galaxy was detected in either one or both of the samples.  This
comparison shows that the published fluxes differ by $\rm \pm 0.1$ dex
owing primarily to differences in the assumptions made for converting 
observed count rates to flux.

Altogether we have 251 galaxies with detected X-ray flux (at the 2.5$\sigma$
level or higher), or 24.7\% of the sample.  Eighty-nine of the detected
galaxies are new, not previously catalogued by F92. Based on the S/N
distribution of ``blank'' fields, as many as 5 of the detections may be
spurious, which yields a true detection rate of 22.6\%.  The majority of
galaxies, 767 in this sample, have only upper limit X-ray values, a result
anticipated at the start of this survey. S/N histograms of the galaxies,
divided into broad Hubble type bins of E+S0+I0, S+Irr and by catalog, show the
expected result that the S-A galaxies, being the nearest, are detected at a
much higher rate than UGC or ESO galaxies.  These data will be used in a
subsequent paper to revisit issues pertaining to X-ray flux generation in
galaxies.

\acknowledgements

This research was supported by a NASA Grant NAG8-665 and 8-734 and by an ASU
Faculty Grant-in-Aid to DB.  APM was supported by NASA JOVE grant NAG8--264.
The hospitality of the Center for Astrophysics is gratefully acknowledged, as
well as the assistance of the Einstein data reduction team. This research has
made use of the NASA/IPAC Extragalactic Database (NED) which is operated by
the Jet Propulsion Laboratory, California Institute of Technology, under
contract with the National Aeronautics and Space Administration. CJF and
WRF acknowledge support from the Smithsonian Institution and the AXAF Science 
Center (NASA contract NAS8--39073).

\newpage

{\bf Figure Captions}

Figure~1. The fraction of galaxies surveyed in each of the three catalogs used
for this survey: a) Shapley-Ames (SA); b) Uppsala General Catalog (UGC); and
c) European Southern Catalog (ESO).  In each case, the galaxies are divided
by Hubble types: E+S0+S0/a galaxies (closed squares); early-type spiral 
galaxies (open circles); late-type spirals + irregulars (open squares) and
multiple/double/peculiar galaxies (open triangles; only in b) and c)).
The fraction of galaxies in our survey in the SA catalog are plotted as a 
function of apparent magnitude.  For both the UGC and ESO catalogs, they are
plotted as a function of log diameter in arcmin.

Figure~2.  The distribution of calculated IPC S/N of the 194 ``blank'' fields,
randomly selected as being free of obvious X-ray emission on the program IPC 
fields, and reduced in the same manner as the program measurments.  The 
distribution is very close to being Gaussian, with a small net offset of
0.095 in S/N.

Figure~3.  a) --- d) The IPC S/N distributions for E+S0 galaxies for
each of the three optical catalogs (a,b,c) plus the three catalogs combined
(d).  In each case, the S/N histogram obtained from the 194 blank fields is
overplotted (dotted line).  Note the good match of the negative tails of the
S/N disributions of the blank fields and program fields for the UGC and 
ESO samples.

Figure~4  a) --- d) IPC S/N distributions for the spiral and irregular 
galaxies in each of the three optical catalogs (a,b,c) plus the three catalogs 
combined (d).  See Figure~3 for details.

Figure~5. F92 X-ray luminosities plotted versus our X-ray lumiosities for 137 
galaxies detected in common between our two surveys.  Closed circles are
X-ray luminosities for E and S0 galaxies; closed squares are early-type 
spirals and closed triangles are late-type spirals and irregulars.

Figure~6. a) The logarithm of the ratio of X-ray luminosity as determined 
in this survey ($\rm L_X(Us)$) to that determined by F92 ($\rm L_X(F92)$),
plotted as a function of $\rm \log L_X(Us)$.  Closed symbols are detections
in both surveys; open symbols are detections by Us but F92 upper limits; 
plus signs, x's and asterisks are detections by F92 but Us upper limits.
Circles and asterisk are E galaxies; squares and plus signs are S0+I0 galaxies;
triangles and crosses are spiral and irregular galaxies.  The dotted lines
mark the median log ratio for E+S0+I0 galaxies (-0.05) and S+I galaxies
(-0.12).  b) The logarithm of measured count rates, Us to F92, now plotted
as a function of RC3 Hubble type number code.  Median values for E+S0+I0 
galaxies (-0.12) and for S+I galaxies (-0.10) are shown as dotted lines.
Here solid squares are mutual detections, open squares are detections by Us 
but F92 upper limits; open triangles are detections by F92 but Us upper limits.


\newpage

\begin{references}

\reference{} Aaronson et al. 1982, \apjs~50, 241

\reference{} Burstein, D., Haynes, M.P. \& Faber, S.M. 1991, Nature 353, 515

\reference{} Burstein, D. \& Heiles, C., 1984, \apjs ~54, 33

\reference{} Burstein, D., Willick, J.A. \& Courteau, S. 1995, in Opacity of 
Spiral Disks, ed. J.I. Davies \& D. Burstein, (Dordrecht: Kluwer), p. 73 

\reference{} Canizares, C. R., Fabbiano, G. \& Trinchieri, G., 1987, \apj ~312, 
503

\reference{} Courteau, S. 1992, Ph.D. Thesis, U.C. Santa Cruz

\reference{} Davies, J.I., Jones, H. \& Trewhella, M. 1995, in Opacity of Spiral
Disks, ed. J.I. Davies \& D. Burstein, (Dordrecht: Kluwer), p. 85

\reference{} de Vaucouleurs, G., de Vaucouleurs, A. \& Corwin, H. G., 1976,
Second Reference Catalogue of Bright Galaxies (Austin: University of Texas
Press); (RC2)

\reference{} de Vaucouleurs, G., de Vaucouleurs, A., Corwin, H. G., Buta, R. J.,
Paturel, G. \& Fouqu\'e, P., 1991, Third Reference Catalogue of Bright
Galaxies (New York: Springer-Verlag); (RC3)

\reference{} Dressel, L. L. \& Condon, J. J., 1976, \apjs ~31, 187

\reference{} Fabbiano, G., Kim, D.-W. \& Trinchieri, G., 1992, \apjs ~80,
531 (F92)

\reference{} Fabbiano, G. \& Trinchieri, G., 1987, \apj ~315, 46

\reference{} Faber, S. M. \& Burstein, D., 1988, in large-scale Motions in the
Universe, eds. V. C. Rubin and G. V. Coyne, Princeton U. Press (Princeton:
NJ), p. 115

\reference{} Faber, S. M., Wegner, G., Burstein, D., Davies, R. L., Dressler,
A., Lynden-Bell, D., \& Terlevich, R. J., 1989, \apjs ~69, 763

\reference{} Forman, W., Jones, C. \& Tucker, W., 1985, \apj~277, 19

\reference{} Giovanelli, R. 1995, in Opacity of Spiral Disks, ed. J.I. Davies
\& D. Burstein, (Dordrecht: Kluwer), p. 127

\reference{} Han, M.S. \& Mould, J.R. 1992, \apj~396. 453

\reference{} Harris et al., 1991, The Einstein IPC Source Catalog

\reference{} Huchra, J., 1976, AJ 81, 952

\reference{} Lauberts, A. \& Valentijn, E. A., 1989, The Surface Photometry 
Catalogue of the ESO-Uppsala Galaxies, (Garching-bei-M\"{u}unchen: European
Southern Observatory)

\reference{} Lauberts, A. 1982, The ESO/Uppsala Survey of the ESO (B) Atlas,
(Garching-bei-M\"{u}unchen: European Southern Observatory) (ESO)

\reference{} Long, K. S. \& Van Speybroeck, L. P., 1981, in Accretion-Driven
X-Ray Sources, ed. W. Lewin \& E. P. J. van den Heuvel (Cambridge: Cambridge
Univ. Press), 117

\reference{} Marshall, F.J. \& Clark, G.W. 1984, ApJ 287, 633

\reference{} Mathewson, D.L., Ford, V.I. \& Buchhorn, M. 1992, \apjs~81, 413

\reference{} Nilson, P., 1973, Uppsala General Catalogue of Galaxies, Nova Acta
R. Soc. Sci. Uppsala, ser. V:A, Vol. 1

\reference{} Roberts, M. S., Hogg, D. E., Bregman, J. N., Forman, W. R. \&
Jones, C., 1991, \apjs ~75, 751

\reference{} Sandage, A. \& Tammann, G. A., 1981, A Revised Shapley-Ames
Catalogue of Bright Galaxies (Washington D.C.: Carnegie Institution of 
Washington)

\reference{} Shapley, H. \& Ames, A. 1932, Ann. Harvard Coll. Obs. 88, No. 2 
(S-A)

\reference{} Stark et al. 1992, ApJS 79, 77 

\reference{} Tormen, G. \& Burstein, D. 1995, \apjs~96, 123 and references 
therein

\reference{} Valentijn, E.A. 1990, Nature 346, 153

\reference{} Willick, J.A., Courteau, S., Faber, S.M., Burstein, D., \& Dekel,
A. 1995, \apj~446, 12

\reference{} Willick, J.A., Courteau, S., Faber, S.M., Burstein, D., Dekel, A.,
\& Kolatt, T. 1996a, \apj~457, 460

\reference{} Willick, J.A., et al, 1996b, \apj, in press.

\reference{} Zwicky, F., Herzog, E., Kowal, C.T., Wild, P. \& Karpowicz, M.
1961, 1963, 1965, 1966, 1968a,b, Catalogue of Galaxies and Clusters of
Galaxies, in six volumes (Pasadena: California Institute of Technology)

\end{references}
\end{document}